\documentclass[prb,twocolumn,showpacs,preprintnumbers,amsmath,amssymb,superscriptaddress]{revtex4}

\usepackage{graphicx}
\usepackage{dcolumn}
\usepackage{bm}
\usepackage{mathrsfs}
\usepackage{subfigure}
\usepackage{natbib}

\begin{document}

\title{Superconducting nanofilms: molecule-like pairing induced by quantum confinement}

\author{Yajiang Chen}
\affiliation{Departement Fysica, Universiteit Antwerpen,
Groenenborgerlaan 171, B-2020 Antwerpen, Belgium}
\author{A. A. Shanenko}
\affiliation{Departement Fysica, Universiteit Antwerpen,
Groenenborgerlaan 171, B-2020 Antwerpen, Belgium}
\author{A. Perali}
\affiliation{School of Pharmacy, Physics Unit, University of
Camerino, I-62032-Camerino, Italy}
\author{F. M. Peeters}
\affiliation{Departement Fysica, Universiteit Antwerpen,
Groenenborgerlaan 171, B-2020 Antwerpen, Belgium}
\email{francois.peeters@ua.ac.be}
\date{\today}
\begin{abstract}
Quantum confinement of the perpendicular motion of electrons in single-crystalline metallic superconducting nanofilms splits the conduction band into a series of single-electron subbands. A distinctive feature of such a nanoscale multi-band superconductor is that the energetic position of each subband can vary significantly with changing nanofilm thickness, substrate material, protection cover and other details of the fabrication process. It can occur that the bottom of one of the available subbands is situated in the vicinity of the Fermi level. We demonstrate that the character of the superconducting pairing in such a subband changes dramatically and exhibits a clear molecule-like trend, which is very similar to the well-known crossover from the Bardeen-Cooper-Schrieffer regime to Bose-Einstein condensation (BCS-BEC) observed in trapped ultracold fermions. For ${\rm Pb}$ nanofilms with thickness of $4$ and $5$ monolayers (${\rm ML}$) this will lead to a spectacular scenario: up to half of all the Cooper pairs nearly collapse, shrinking in the lateral size (parallel to the nanofilm) down to a few nanometers. As a result, the superconducting condensate will be a coherent mixture of almost molecule-like fermionic pairs with ordinary, extended Cooper pairs.
\end{abstract}

 \maketitle

\section{Introduction}

Advances in self-assembly and nanofabrication resulted recently in the thinnest superconductors, i.e., single-crystalline metallic nanofilms with atomically uniform thickness down to a few monolayers, see, e.g., Refs.~\onlinecite{eom,cren,qin,zhang,weit}. They are characterized by an extraordinary purity and, so, the only issue of disorder in the case of interest is due to the interface between the substrate and the nanofilm. Such an interface has an effect on the reflection of electron waves (it is not perfectly specular). Nevertheless, there are clear signatures of the formation of discrete levels for the perpendicular electron motion [i.e., quantum-well states (QWS)] in tunneling spectra.~\cite{eom,qin} Pairing correlations appear to be very robust in single-crystalline nanofilms so that even ${\rm Pb}$ and ${\rm In}$ films with
thickness of $1\,{\rm ML}$ were found not to exhibit any considerable signs of
degradation of superconductivity due to fluctuations.~\cite{zhang} Experimental results for the temperature dependence of the pairing gap show almost no deviations from the ordinary BCS picture, see, e.g., Refs.~\onlinecite{eom,qin,zhang}. However, it is worth noting that, for some samples, there appear signatures of suppression of density of states (DOS) in the tunneling spectra at low voltage above $T_c$, which resembles, to some extent, the pseudogap physics of high-$T_c$ superconductors.~\cite{zhang,wang}

Another recent experimental breakthrough concerns the iron-pnictides (for a review see, e.g., Ref.~\onlinecite{pnic}), i.e., a new generation of multi-band superconductors specified by the interplay of different band-dependent superconducting condensates. This renewed the interest in multi-band superconductivity. In particular, one of the most intriguing points is a competition of the characteristic lengths of the different condensates~\cite{bab,mosch,kogan,shan} and a possible contribution of such a competition to new phenomena, including recently observed and strongly debated unconventional patterns of the vortex distribution in magnesium-diboride.~\cite{mosch}

The question arises if there is any relation between bulk multi-band superconductors and high-quality superconducting nanofilms? In fact, they are similar in many important respects. As already mentioned above, tunneling experiments demonstrate that the conduction band in metallic nanofilms splits up into a series of subbands in such a way that the bottom of each subband is at the energy position of the corresponding QWS.~\cite{eom,qin,wang} For example, ${\rm Pb}$ nanofilms with thickness $4$-$5\,{\rm ML}$ are two-band(subband) superconductors.~\cite{eom,qin} In those metallic superconducting nanofilms the
energetic position of each subband with respect to the Fermi level $E_F$ changes with thickness, i.e., the discrete perpendicular levels scale as $1/d^2$, with $d$ the thickness of the nanofilm while the accompanying variation of $E_F$ is almost
insignificant.~\cite{note} Additional reasons for shifting in energy of the subband bottoms as measured from $E_F$ are due to fabrication circumstances: the presence or absence of a protective cover, an effect of the wetting layer, use of different substrates etc., which can significantly change the behavior of the
single-electron wave functions at the interface (and, so, the QWS energy). Thus, the relevant microscopic parameters of the different subbands, e.g., DOS and the Fermi velocity (in the parallel direction), are not fixed and can significantly vary from sample to sample even for the same thickness. It is possible to expect that these parameters can, in principle, be tailored in future experiments. Therefore, it is timely to investigate in more detail the properties of the superconducting state in metallic single-crystalline nanofilms in the case when the bottom of one of the relevant single-electron subbands approaches the Fermi level. Below we show that the pairing in such a subband exhibits a clear molecule-like trend, which is very similar to the BCS-BEC crossover investigated at length in cold atomic gases, see, e.g., Ref.~\onlinecite{bloch}. In particular, we predict that striking results can be obtained for ${\rm Pb}$ nanofilms with thicknesses $4$-$5\,{\rm ML}$, where up to half of all the Cooper pairs nearly collapse, i.e., shrinking in the lateral size (parallel to the nanofilm) down to a few nanometers. Our finding significantly compliments the recent first observation of the BCS-BEC crossover in a solid-state material, i.e., in one of the available subbands in a multiband superconducting iron-chalcogenide ${\rm Fe Se_xTe_{1-x}}$, see Ref.~\onlinecite{lub}.

The present paper is organized as follows. In Sec.~\ref{one-band} we address the criterion of the BCS-BEC crossover-like behavior in a given single-electron subband. We show that a condensed pair of electrons significantly shrinks in its lateral size (parallel to the nanofilm) when the ratio of the parallel kinetic energy of electrons to the absolute value of the potential energy becomes smaller than one, i.e., when the subband bottom approaches the Fermi level. In Sec.~\ref{two-band} we investigate $4$ and $5\,{\rm ML}$ thick lead
nanofilms, where only two single-electron subbands are occupied, i.e., we have a coherent mixture of two condensates, and one of them is specified by an extremely small characteristic length. Our conclusions are given in Sec.~\ref{con}.

\section{BCS-to-BEC crossover driven by quantum-size effects}
\label{one-band}

It is well-known that the governing parameter for the BCS-BEC crossover in the system with fermionic pairing correlations is the ratio of the relevant kinetic energy $K$ to the absolute value of the potential energy $U$. When $K/|U| > 1$, we are in the BCS regime with loosely bound and rather extensive Cooper pairs (the BCS limit corresponds to $K/|U| \to \infty$). On the opposite side of the crossover, when $K/|U| \ll 1$, molecule-like bound pairs appear with the BEC limit $K/|U| \to 0$. Here it is worth noting that an extensive analysis of the behavior of the kinetic and potential energy through the BCS-BEC crossover for the attractive Hubbard model~\cite{toschi} gave $K/|U|\approx 2.2$ at the BCS side, while approaching the BEC regime it resulted in $K/|U|\approx 0.03$. Interestingly, the right balance between the strength of the kinetic and interaction energy has been considered as an essential feature of high-$T_c$ superconductivity.~\cite{basov}

Keeping in mind the criterion based on the ratio $K/|U|$, let us examine what happens with a Cooper pair in a given subband when changing the position of its bottom with respect to $E_F$ in single-crystalline nanofilms. If the bottom of the subband is situated far below the Fermi level, the mean kinetic energy of  electrons in this subband comes mostly from the parallel motion, i.e., $K_{\perp} \ll K_{||} \sim E_F$. In this case $K_{||}/|U|\sim E_F/|U| \gg 1$. However, when the subband bottom (or the corresponding QWS) approaches the Fermi surface, we have $K_{\perp} \to E_F$ and, so, $K_{||}$ becomes much smaller than $E_F$. The parallel motion of electrons is reduced, and such a redistribution of the kinetic energy between the perpendicular and parallel spatial degrees of freedom leads to a significant decrease in the ratio $K_{||}/|U|$. Furthermore, when the subband bottom goes above the Fermi level (such a subband still makes a contribution
unless its bottom is above $E_F + \hbar\omega_D$, with $\omega_D$ the Debye frequency), $K_{||} \to 0$ and, hence, $K_{||}/|U|$ becomes extremely small, i.e., $K_{||}/|U| \ll 1$. Therefore, when assuming that the ratio $K_{||}/|U|$ controls the lateral size of a condensed fermionic pair associated with the corresponding subband in the nanofilm, we expect from the above arguments a significant reduction in the lateral size of the Cooper pairs.

To go into more detail, let us consider a superconducting nanoslab in the clean limit, with perfectly specular reflection at the boundaries. Possible effects of the scattering of electrons at the interface between a substrate and the nanofilm will be discussed later. The translational invariance in the perpendicular direction is broken due to quantum confinement and, so, we deal with a spatially nonuniform order parameter $\Delta({\bf r})$ problem. As a consequence, the ordinary BCS self-consistent equation should be abandoned in favor of a more elaborate analysis based on the Bogoliubov-de Gennes (BdG) equations (or, equally,
on the Gor'kov Green's function formalism). As shown in Ref.~\onlinecite{pieri}, the BdG equations are appropriate to describe the BCS-BEC crossover in spatially nonuniform fermionic systems at nearly zero temperatures. In particular, it has been shown in Ref.~\onlinecite{pieri} that the BdG equations reproduce the Gross-Pitaevskii equation for the condensate wave function at the BEC side of the crossover. Therefore, if our reasoning based on $K_{||}/|U|$ is correct, we can expect that the spatial profile of the wave function of a condensed electronic pair will change (dramatically) crossing over from the BCS regime~(i.e., $K_{||}/|U| \gg 1$) to the BEC regime (i.e., $K_{||}/|U| \ll 1$). In the BCS regime this pair-wave function will have many oscillations with the period of the Fermi wavelength $\lambda_F$ and will decay over a significantly larger distance (as compared to $\lambda_F$) determined by the size of the extended Cooper pairs. Approaching the BEC regime, the pair-wave function will have only some insignificant residual oscillations~(e.g., due to the presence of the ultraviolet cut-off) and will be concentrated at short separations between electrons associated with the size of a molecule-like pair (it is close to or smaller than $\lambda_F$). For a superconducting slab the BdG equations can be written in the form
\begin{subequations}\label{bdg}
\begin{align}
&\varepsilon_{i,{\bf k}}|u_{i,{\bf k}}\rangle=\hat{H}_e|u_{i,
{\bf k}}\rangle + {\hat{\Delta}}|v_{i,{\bf k}}\rangle,
\label{bdgA} \\
&\varepsilon_{i,{\bf k}}|v_{i,{\bf k}}\rangle={\hat{\Delta}}^{
\ast}|u_{i,{\bf k}}\rangle - \hat{H}_e^{\ast}|v_{i,{\bf k}}\rangle,
\label{bdgB}
\end{align}
\end{subequations}
where $i$ is the quantum number associated with the quantum-confined motion in the $z$ direction, ${\bf k}$ is the wavevector of the quasi-free motion of electrons in the direction parallel to the nanofilm ($x$ and $y$ directions); $|u_{i,{\bf k}}\rangle$ and $|v_{i,{\bf k}}\rangle$ are the particle-like and hole-like ket vectors; $\varepsilon_{i,{\bf k}}$ is the quasiparticle energy. In addition, $\hat{\Delta} = \Delta(\hat{\bf r})$, with $\hat{\bf r}$ the electron position operator, and $\hat{H}_e=\hat{H}^\ast_e$~($\ast$ stands for the complex conjugate) is the single-electron Hamiltonian~(measured from the Fermi level $E_F$)
\begin{equation}
\hat{H}_e = \frac{\hat{\bf p}^2}{2m_e} + V_{\rm conf}(\hat{\bf r}) - E_F,
\label{He}
\end{equation}
with $m_e$ the band mass set to the free electron mass and $V_{\rm conf}({\bf r})$ the confining potential. Equations (\ref{bdg}) are solved together with the self-consistency relation given by
\begin{equation}
\Delta({\bf r}) = g\sum\limits_{i,{\bf k}} \langle {\bf r}|u_{i,{\bf k}}\rangle \langle v_{i,{\bf k}}|{\bf r}\rangle,
\label{self}
\end{equation}
where $g$ is the coupling constant, and we limit ourselves to zero temperature. We ignore pairing of electrons between different subbands, which is justified when the intersubband energy spacing, i.e., $\delta \sim  \frac{\hbar^2}{2m} \frac{\pi^2}{ d^2}$, is significantly larger than the pairing energy (it is always true for ultrathin films). In this case it is possible to take into account only the pairing of the time reversed states, which means that (see, e.g., Ref.~\onlinecite{shan1})
\begin{equation}
|u_{i,{\bf k}}\rangle={\cal U}_{i,{\bf k}}|\xi_{i,{\bf k}}\rangle,
\quad |v_{i,{\bf k}}\rangle={\cal V}_{i,{\bf k}}|\xi_{i,{\bf
k}}\rangle, \label{tr}
\end{equation}
where $\hat{H}_e|\xi_{i,{\bf k}}\rangle=\xi_{i,{\bf k}}|\xi_{i,{\bf k}}\rangle$, with $\xi_{i,{\bf k}}$ the single-electron energy measured from the Fermi level, and the factors ${\cal U}_{i,{\bf k}}$ and ${\cal V}_{i,{\bf k}}$ are chosen real,
together with the order parameter. Based on Eqs.~(\ref{bdg}), (\ref{self}) and (\ref{tr}), one can find
\begin{equation}
{\cal U}^2_{i,{\bf k}}=\frac{1}{2}\Big[1+ \frac{\xi_{i,{\bf k}}}{
\varepsilon_{i,{\bf k}}}\Big], \quad {\cal V}^2_{i,{\bf
k}}=\frac{1}{2} \Big[1-\frac{\xi_{i,{\bf k}}}{\varepsilon_{i,{\bf
k}}}\Big], \label{tr1}
\end{equation}
with $\varepsilon_{i,{\bf k}}=\sqrt{\xi^2_{i,{\bf k}}+\Delta^2_i}$, where the subband-dependent gap $\Delta_i=\langle \xi_{i,{\bf k}}|\hat{\Delta}|\xi_{i,{\bf k}}\rangle$ obeys the following BCS-like self-consistency equation:
\begin{equation}
\Delta_{i'} = \sum_{i,{\bf k}} \Phi_{i',i} \frac{\Delta_i}{2
\varepsilon_{i,{\bf k}}}, \label{tr2}
\end{equation}
with $\Phi_{i',i}=g\int {\rm d}^2r |\langle {\bf r}|\xi_{i',{\bf k}}\rangle|^2 |\langle {\bf r}|\xi_{i,{\bf k}}\rangle|^2$. As usual, to avoid the ultraviolet divergence, the sum in Eq.~(\ref{tr2})~[and in Eq.~(\ref{self})] runs over the states with $|\xi_{i,{\bf k}}| < \hbar\omega_D$, where $\omega_D$ stands for the Debye frequency. For ${\rm Pb}$ we adopt $\hbar\omega_D =8.27\,{\rm meV}$, see Ref.~\onlinecite{fett}.

To check our arguments about the influence of the ratio $K_{||}/U$ on the lateral subband-dependent size of the Cooper pairs, we need to study the "wave function"
of a condensed fermionic pair, i.e., the anomalous average of the field operators $\langle {\hat \psi}_{\uparrow} ({\bf r}) {\hat \psi}_{\downarrow}({\bf r}')\rangle$. More precisely, the anomalous average is proportional to the bound-like eigenstate of the two-particle density matrix, which can be safely interpreted as the wave function of a condensed fermionic pair, see, e.g., Refs.~\onlinecite{bog} and \onlinecite{cherny}. This quantity is related to the particle-like and  hole-like vectors of the BdG equations through the Bogoliubov canonical transformation
\begin{subequations}\label{bct}
\begin{align}
&{\hat \psi}_{\uparrow}({\bf r})=\sum\limits_{i,{\bf k}} \Big[
\langle {\bf r}|u_{i,{\bf k}}\rangle\,\gamma_{i,{\bf k},\uparrow} -
\langle v_{i,{\bf k}}|{\bf r}\rangle\, \gamma^{\dagger}_{i,-
{\bf k},\downarrow}\Big],\label{bctA}\\
&{\hat \psi}_{\downarrow}({\bf r})=\sum\limits_{i,{\bf k}}
\Big[\langle {\bf r}|u_{i,{\bf k}}\rangle\,\gamma_{i,-{\bf k},
\downarrow} + \langle v_{i,{\bf k}}|{\bf r}\rangle\, \gamma^{
\dagger}_{i,{\bf k},\uparrow}\Big], \label{bctB}
\end{align}
\end{subequations}
where $\gamma^{\dagger}$ and $\gamma$ are the quasiparticle (bogolon) operators. From Eq.~(\ref{bct}) we find (for $T=0$)
\begin{equation}
\langle {\hat \psi}_{\uparrow}({\bf r}) {\hat \psi}_{\downarrow}
({\bf r}')\rangle= \sum\limits_{i,{\bf k}} {\cal U}_{i,{\bf k}}
{\cal V}_{i,{\bf k}}\; \langle {\bf r}|\xi_{i,{\bf
k}}\rangle\langle\xi_{i,{\bf k}}|{\bf r}'\rangle, \label{wf}
\end{equation}
which can be rearranged to
\begin{equation}
\langle {\hat \psi}_{\uparrow}({\bf r}) {\hat \psi}_{\downarrow}
({\bf r}')\rangle= \sum\limits_i \frac{\Delta_i}{2}\,\varphi_i(z) \varphi^{\ast}_i(z')\,\eta_i(x-x',z-z'),
\label{wf1}
\end{equation}
where $\varphi_i(z)$ is the single-electron wave function associated with the $i-$th QWS, and $\eta_i(x-x',y-y') = \eta_i(\rho)$~[with $\rho =\sqrt{(x-x')^2+ (y-y')^2}$] controls the decay of the fermionic-pair "wave function" in the direction parallel to the nanofilm. This quantity can be represented in the form
\begin{equation}
\eta_i(\rho)=\int\!\!\frac{{\rm d}k}{2\pi}\,k\;\theta(\hbar\omega_D
-|\xi_{i,{\bf k}}|)\;\frac{J_0(k\rho)}{\sqrt{\xi^2_{i,{\bf k}}
+\Delta^2_i}}, \label{wfxy}
\end{equation}
with $k=\sqrt{k^2_x+k^2_y}$ and $J_0$ the Bessel function of the first kind of order $0$. Let us here introduce the QWS energy $\varkappa_i$, which makes it possible to write $\xi_{i,{\bf k}}=\frac{\hbar^2}{2m}(k^2_x + k^2_y) - \mu_i$,
with $\mu_i =E_F-\varkappa_i$.
%
\begin{figure*}
\includegraphics[width=1.0\linewidth]{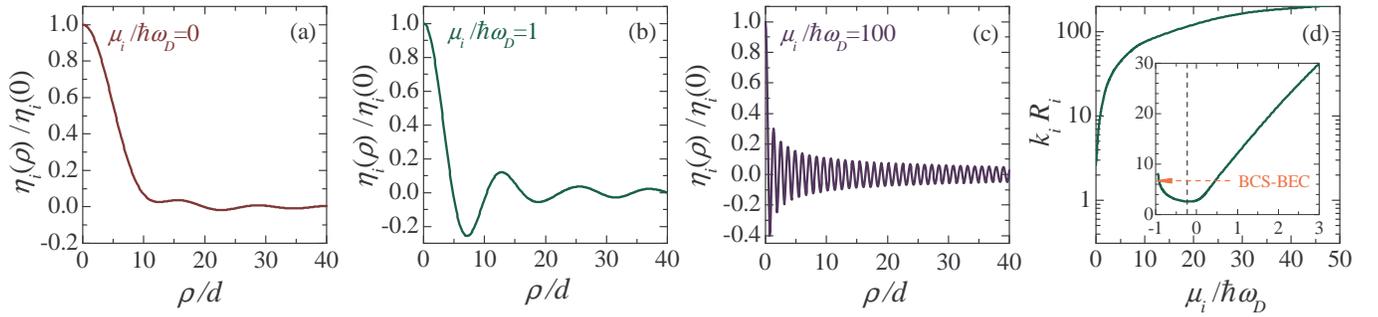}
\caption{(Color online) (a,b,c) The subband-dependent pair "wave function" $\eta_i(\rho)$ versus $\rho$ for different values of $\mu_i$, i.e., $\mu_i/\hbar\omega_D=0$~(a), $\mu_i/ \hbar\omega_D=1$~(b), and $\mu_i/\hbar\omega_D=100$~(c). Panel (d) shows how the product $k_iR_i$ depends on the ratio $\mu_i/\hbar\omega_D$~($1/k_i$ is the measure for the mean distance
between electrons in a given subband and $R_i$ stands for the pair size). The inset in (d) is a room around $\mu_i=0$, which shows details of a drop in the product $k_iR_i$ when the bottom of the corresponding single-electron subband approaches the Fermi level (the BCS-BEC crossover is reached below the dashed line, i.e., $k_iR_i < 2\pi$).}
\label{fig1}
\end{figure*}
%
The asymptote for $\eta_i(\rho)$ at large $\rho$ can be calculated analytically
in several interesting cases. When the QWS level (i.e., the subband bottom) is situated far below the Fermi level, i.e., $\mu_i > 0$ and $\mu_i \gg \Delta_i$,
we find
\begin{equation}
\eta_i(\rho) \simeq \frac{m}{\pi \hbar^2}
J_0(k_i\rho)\,K_0(\rho/R^{(1)}_i)\quad \big[\rho/R^{(1)}_i \gtrsim
1\big], \label{BCS}
\end{equation}
where $K_0$ is the Macdonald function and $R^{(1)}_i=\hbar v_i/\Delta_i$, with $v_i= \hbar k_i/m$ the subband-dependent Fermi velocity (for parallel motion). This is nothing more but the ordinary BCS behavior: first, there are fast oscillations with period of the subband-dependent Fermi wavelength $\lambda_i =2\pi/k_i$~[this comes from $J_0(k_i\rho)$]; second, we obtain the exponential overall decay governed by $R^{(1)}_i$, as seen from the asymptote of $K_0(\rho/R^{(1)}_i)$.

When QWS touches the Fermi level, i.e., $\mu_i=0$, Eq.~(\ref{BCS}) does not hold any more and the behavior of $\eta_i(\rho)$ changes dramatically. In this case, assuming $\rho \to \infty$, we arrive at
\begin{equation}
\eta_i(\rho) \simeq\frac{m}{\pi\hbar^2} J_0(\rho/R^{(2)}_i)\,K_0(\rho/ R^{(2)}_i) \quad \big[\rho/R^{(2)}_i\gtrsim 1\big], \label{cross}
\end{equation}
with $R^{(2)}_i=\hbar/\sqrt{m\Delta_i}$. As seen, the characteristic length controlling the decay of the pair "wave function" in Eq.~(\ref{cross}) is completely different from that in Eq.~(\ref{BCS}). Moreover, $R^{(2)}_i$ is significantly smaller than $R^{(1)}_i$. As $R^{(1)}_i$ can be well approximated by $\hbar v_F/\Delta_i$, with $v_F=\sqrt{2E_F/m}$, then we obtain
\begin{equation}
R^{(1)}_i/R^{(2)}_i \propto \sqrt{E_F/\Delta_i}.
\label{ratio}
\end{equation}
In addition, the fast oscillations present in Eq.~(\ref{BCS}) disappear in Eq.~(\ref{cross}). Instead, we obtain a rather slowly oscillating factor $J_0 (\rho/R^{(2)}_i)$ but the role of these oscillations is almost negligible because they can only manifest themselves for separations of electrons in a condensed pair such that $\eta_i(\rho)$ almost approaches zero. So, our expectations based on the behavior of the parameter $K_{||}/|U|$ are relevant: when the subband bottom touches the Fermi level, the Cooper pairs associated with this subband nearly collapse (as compared to typical values of the zero-temperature BCS coherence length) in the lateral direction, shrinking by a factor of $\sqrt{E_F/\Delta_i} \sim 10^2$. Such shrinking continues when $\mu_i$ crosses zero and becomes negative. In particular, when $\hbar\omega_D\gg|\mu_i|\gg \Delta_i$,  Eq.~(\ref{wfxy}) results in
\begin{equation}
\eta_i(\rho) \simeq \frac{m}{\pi\hbar^2}\,K_0(\rho/R^{(3)}_i)\quad
\big[\rho/R^{(3)}_i \gtrsim 1\big], \label{BEC}
\end{equation}
with $R^{(3)}_i =\hbar/\sqrt{2m|\mu_i|}$. As $R^{(3)}_i \approx
\sqrt{\hbar /(2m\omega_D)}$, we obtain
\begin{equation}
R^{(1)}_i/R^{(3)}_i \propto \sqrt{E_F \hbar\omega_D}/\Delta_i,
\label{ratio1}
\end{equation}
which can be larger than the ratio in Eq.~(\ref{ratio}) by a factor of $\sqrt{\hbar \omega_D/\Delta_i}$, which is about $3-10$ for conventional superconductors. Thus, based on Eqs.~(\ref{ratio}) and (\ref{ratio1}), we find a giant drop, from microns to nanometers, in the lateral size of a condensed fermionic pair associated with the single-electron subband whose bottom approaches $E_F$. This molecule-like trend in the reconstruction of the electronic pairing is very similar to the one predicted previously for high-quality superconducting nanowires.~\cite{shan1}

More details about $\eta_i(\rho)$ can be obtained by numerically calculating the integral in Eq.~(\ref{wfxy}). Using $\Delta_i =1.08\,{\rm meV}$, which results from tunneling measurements of $4\,{\rm ML}$ thick lead nanofilms,~\cite{eom,qin} we investigate how the total profile of $\eta_i$ evolves with changing $\mu_i$, in addition to the asymptotic behavior discussed in the previous paragraph. Figure~\ref{fig1} shows numerical results for $\eta_i(\rho)$ for different values of $\mu_i$, i.e., $\mu_i/\hbar\omega_D=0$~(a), $\mu_i/\hbar\omega_D=1$~(b), and $\mu_i/\hbar\omega_D=100$~(c). As seen, the numerical results fully support the analytical results given by Eqs.~(\ref{BCS}), (\ref{cross}) and (\ref{BEC}). In particular, we obtain the typical BCS picture of anomalous correlations when the subband bottom is situated far below $E_F$~[see panel (c)] but we approach a
molecule-like character of the spatial distribution of electrons in a condensed pair for $\mu_i \to 0$~[see panel (a)]. Fast oscillations in $\eta_i(\rho)$ at Fig.~\ref{fig1}(c) are first converted into slow oscillations shown in panel (b)~[here $\lambda_i$ significantly increases as compared to panel (c)] and, then, are almost washed out in Fig.~\ref{fig1}(a). Only small residual oscillations can be observed in panel (a), and the period of these oscillations is about the decay radius for $\eta_i(\rho)$, which is in agreement with Eq.~(\ref{cross}).
%
\begin{figure}
\includegraphics[width=0.6\linewidth]{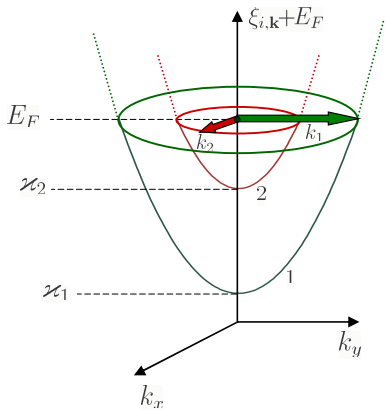}
\caption{(Color online) Sketch of the single-electron energy spectrum
$\xi_{i,{\bf k}}+E_F$, with ${\bf k}=\{k_x,k_y\}$, for the lower~(1) and upper~(2) subbands in $4$-$5\,{\rm ML}$ lead single-crystalline nanofilms. Two thick
arrows show the subband-dependent Fermi wavenumbers.}
\label{fig2}
\end{figure}
%
To further highlight similarities with the ordinary BCS-BEC crossover driven by the Feshbach resonance in ultracold Fermi gases, we show in Fig.~\ref{fig1}(d) how the product $k_iR_i$ depends on $\mu_i$, where $R_i$ is the mean square radius calculated with the "wave function" $\eta_i(\rho)$ and $k_i = (3\pi^2 n_i)^{1/3}$, with $n_i$ the mean electron density in subband $i$. We note that here $k_i$ is introduced to measure the mean distance $\approx\lambda_i=2\pi/k_i$ between the electrons in the $i$-th subband, notwithstanding the presence/absence of the real Fermi motion and the position of the subband bottom with respect to $E_F$. When the bottom of a subband is far below the Fermi level this definition of $k_i$~(and of $\lambda_i$) recovers the definition used in Eq.~(\ref{BCS}). Following the results of Ref.~\onlinecite{pis}, we can expect that the BCS-BEC crossover is approached in a given subband when $1/\pi < k_iR_i< 2\pi$. In other words, the size of a condensed fermionic pair $R_i$ becomes smaller than the mean distance between electrons in the corresponding subband. Having at our disposal this inequality, we can learn from Fig.~\ref{fig1}(d), that the BCS-BEC crossover is definitely approached when the subband bottom approaches $E_F$, i.e., when $|\mu_i|/\hbar\omega_D \lesssim 1$. Here it is worth noting that, as seen from the inset in panel (d), the product $k_iR_i$ slightly increases with decreasing $\mu_i$ for $\mu_i < -0.2\hbar\omega_D$. This feature appears due to the presence of the ultraviolet cut-off and demonstrates that the regime of Eq.~(\ref{BEC}) is not reached in our calculations. The point is that Eq.~(\ref{BEC}) essentially assumes that $\hbar\omega_D \gg |\mu_i| \gg \Delta_i$ but this can not be realized for the parameters used to calculate the data of Fig.~\ref{fig1}, i.e., for $\hbar\omega_D =8.27\,{\rm meV}$ and $\Delta_i=1.08\,{\rm meV}$. It is also of importance to note here that despite a significant shrinking in their lateral size down to $\lambda_i$, the condensed pairs of electrons do not suffer from the Coulomb repulsive effects. In metals such effects can be expected only for the separations smaller than $0.3$-$0.4\,{\rm nm}$ whereas in the subband whose bottom is in the vicinity of the Fermi level $\lambda_i \sim 5$-$10\,{\rm nm}$, i.e., the mean density of electrons in the subband whose bottom approaches $E_F$ is significantly smaller than the total mean electron density. We stress that this does not prevent this subband from making a significant contribution to the coherent phenomena, see Sec.~\ref{two-band}.

Based on the results of this section, we can conclude that the redistribution of the kinetic energy between the perpendicular and parallel spatial degrees of freedom leads to a significant reconstruction of the internal distribution of electrons in a condensed pair so that we find a clear molecule-like trend in the
pairing. This behavior is found for the pair condensate in a subband whose bottom approaches the Fermi level. In total, we have a coherent mixture of different  subband-dependent condensates. So, the question arises what are the consequences of this mixture in the presence of the BCS-BEC crossover-like behavior in one of the available subbands. This point is addressed in the next section.

\section{Two condensates and two characteristic lengths}
\label{two-band}

\begin{figure}
\includegraphics[width=1.0\linewidth]{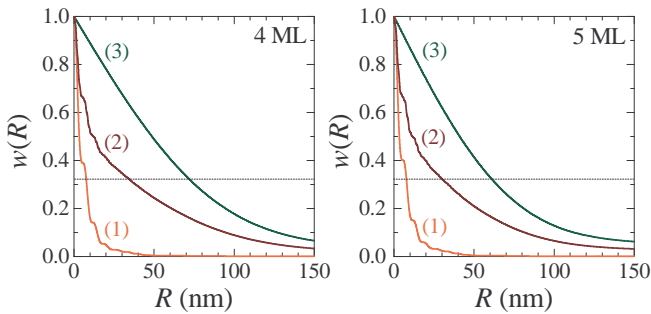}
\caption{(Color online) The probability of finding a Cooper pair with lateral size larger than $R$ for ${\rm Pb}$ superconducting nanofilms with thickness $4\,{\rm ML}$~(the left panel) and $5\,{\rm ML}$~(the right panel). The results for a condensed fermionic pair in the upper subband are given by curve 1; the total probability for the both contributing subbands is represented by curve 2; the results for the Cooper pairs associated with the lower subband are given by 3. The Fermi level is adjusted so that the bottom of the upper of the two available subbands is at the energy $E_F -\hbar\omega_D$. The dotted line in both panels correspond to $w=0.32$.} \label{fig3}
\end{figure}

In the previous section we discussed the results for the fermionic condensate
associated with one of the single-electron subbands appearing due to quantization of the perpendicular electron motion. However, due to the presence of multiple subbands we have a coherent mixture of subband-dependent fermionic condensates each making a different contribution to the total condensate and superconducting characteristics. In nanofilms, due to the quasi-2D character of the electron motion, such contributions are, as a rule, almost equal to each other, except in the case when the subband bottom becomes significantly higher than $E_F - \hbar\omega_D$. In particular, when it approaches the level $E_F+\hbar\omega_D$, the corresponding subband stops to contribute due to the ultraviolet cut-off. The most interesting choice is realized when there are only two relevant subbands,~\cite{note} which is the case for, e.g., lead nanofilms with thicknesses $4$ and $5\,{\rm ML}$~(see the sketch given in Fig.~\ref{fig2}). The optimal variant of tuning the subband positions is when the bottom of the upper subband is situated at $E_F - \hbar\omega_D$. First, the contribution of such a subband to the superconducting characteristics is almost the same as that of the lower subband with the bottom far below $E_F$. Second, the lateral size of the condensed fermionic pairs associated with the upper subband in this case only slightly increases as compared to $\mu_i=0$, see Fig.~\ref{fig1}(d). Our numerical results for this variant are shown in Fig.~\ref{fig3} for lead nanofilms with thicknesses $4\,{\rm ML}$~(the left panel) and $5\,{\rm ML}$~(the right panel). $1\,{\rm ML}$ corresponds to $0.286\,{\rm nm}$ in our calculations.~\cite{wei}
Here the quantity
\begin{align}
w(R)=1\;-\;\frac{\iint\limits_{\rho < R}{\rm d}^3r{\rm d}^3r'\langle {\hat \psi}_{ \uparrow}({\bf r}) {\hat\psi}_{\downarrow} ({\bf r}')\rangle}{\iint\!{\rm d}^3r{\rm d}^3r'\langle {\hat \psi}_{\uparrow}({\bf r}) {\hat \psi}_{\downarrow}({\bf r}')
\rangle},
\label{w}
\end{align}
i.e., the probability of finding a Cooper pair with the lateral size larger than $R$, is plotted versus $R$. In both panels of Fig.~\ref{fig3} the different curves are for: (1), the results for the upper subband, whose bottom touches $E_F-\hbar\omega_D$~[when keeping only its contribution to Eq.~(\ref{wf1})]; (2), the total results of both contributing subbands; and (3), the data for the lower single-electron subband whose bottom is far below $E_F$. Here the coupling constant $g$ is chosen such that the experimental critical temperature $T_c$, as reported in Ref.~\onlinecite{qin}, is obtained. For illustrative purposes, the confining interaction was taken as zero inside the nanofilm and infinite outside. The Fermi level $E_F$ is adjusted so that to get the second QWS at $E_F -\hbar\omega_D$, i.e., $E_F=\varkappa_2 + \hbar\omega_D$. As seen from Fig.~\ref{fig3}, the results for $4\,{\rm ML}$ and $5\,{\rm ML}$ are quite close to each other. If we adopt that the lateral diameter of the fermionic pair is defined by $w(R)=0.32$~(recall that for the normal distribution $1-\int_{-\sigma}^{+\sigma}{\rm  d}\tau\rho_{norm}(\tau)=0.32$), then we find that the Cooper-pair radius associated with the lower subband is $R_{\rm low}=72\,{\rm nm}$ for $4\,{\rm ML}$ and $62\,{\rm nm}$ for $5\,{\rm ML}$. The minor difference here is due to slightly different Fermi velocities (in the parallel direction) and critical temperatures. As for the pair radius for the upper subband, we obtain $R_{\rm up}=7.6\,{\rm nm}$ for (a) and $R_{\rm up}=7.8\,{\rm nm}$ for (b). These results are well explained by the trend given by Eq.~(\ref{cross}), i.e., $R_{\rm up} \propto \hbar/ \sqrt{mk_BT_c}$, with $T_c=6.7\,{\rm K}$ for $4\,{\rm ML}$ and $6.3\,{\rm K}$ for $5\,{\rm ML}$~(see Ref.~\onlinecite{qin}). For the total radius we have for curve (2) $R_{\rm tot}=34\,{\rm nm}$ for panel (a) and $R_{\rm tot}=32\,{\rm nm}$ for panel (b). The drop of $R_{\rm tot}$ as compared to the bulk-like result for the lower subband is due to the contribution of the extremely small pairs associated with the upper subband. It is worth noting that the introduction of the characteristic pair size for a coherent mixture of two different condensates is somewhat conventional. As seen from Fig.~\ref{fig3}, curve (2) exhibits a bimodal behavior with a clear crossover from the short-range regime governed by the upper band to the long-range one controlled by the lower band. Such a behavior can not be, of course, accurately specified by one characteristic length. This is in agreement with the recent analytical results for the extended Ginzburg-Landau formalism for multiband superconductors~\cite{shan} and with numerical investigations of the coherent properties of ${\rm MgB_2}$, see, e.g., Ref.~\onlinecite{kosh}. Thus, $R_{\rm tot}$ measures the short-to-long crossover in $w(R)$ rather than the size of a condensed fermionic pair. Nevertheless, the probability to pick up a condensed pair with lateral size smaller than $R_{\rm tot}$ is about $0.68$. Moreover, as $w(R)$ decreases relatively slow around $R_{\rm tot}$, the probability to find a condensed pair with size smaller than $10\,{\rm nm}$ is about $0.5$: almost half of the condensed fermionic pairs are smaller than $10\,{\rm nm}$ in lateral size.

%
%

What are possible experimental evidences of such shrinking of the Cooper pairs due to the BCS-BEC crossover-like behavior in single-crystalline nanofilms? First of all, this is a competition of different characteristic spatial lengths of two condensates. It was recently shown that such a competition can lead to the appearance of long-range attraction between vortices in a two-band superconductor,
see Ref.~\onlinecite{sil}. In this case the Abrikosov lattice of vortices formed in the perpendicular magnetic field will melt, with possible formation of stripes and clusters of vortices.~\cite{bab,mosch} Second, the profile of the order parameter in the core of the vortex will be sensitive to the presence of the condensate with extremely small spatial coherence length, which can be probed via the scanning tunneling microscopy, see, e.g., Ref.~\onlinecite{cren}. Third, we can also expect
significant proliferation of pair fluctuations in the presence of a single-electron subband with the bottom next to $E_F$. Such fluctuations will result in the suppression of the DOS at the Fermi level above the critical temperature~\cite{perali}, i.e., the pseudogap behavior. However, it is difficult to say whether or not these pairing fluctuations make a contribution to the pseudogap-like behavior revealed recently in single-crystalline lead nanofilms~\cite{wang}: other mechanisms can also be involved~(see discussion in Ref.~\onlinecite{wang}) and, so, a more involved analysis is needed. In addition, Josephson physics and Andreev states will be also very sensitive to the appearance of the molecule-like trend in pairing and to the competition of different length-scales. In particular, these important things can be probed with hybrid superconducting devices similar to, e. g., a carbon nanotube joining a superconducting loop.~\cite{car} In our case the single-crystalline metallic nanofilm is supposed to transmit the current between superconductors over  mesoscopic distances (to keep the nanofilm in the normal state it is necessary to have the critical temperature of leads higher than that of the nanofilm).

\section{Conclusion}
\label{con}

In conclusion, we demonstrated that quantum confinement of the perpendicular motion of electrons in single-crystalline metallic superconducting nanofilms leads to the formation of a nanoscale multi-band(subband) superconductor. The position in energy of each subband can vary significantly depending on the fabrication
process, and it is possible that the bottom of one of such subbands is situated in the vicinity of the Fermi level. We showed that the character of the  superconducting pairing in such a subband changes dramatically and exhibits a clear molecule-like trend, which is very similar to the BCS-BEC crossover but now driven by the perpendicular size-quantization.

Though we have considered the system in the clean limit, our results will also be relevant in the presence of a moderate disorder. The main issue for disorder in high-quality metallic superconducting nanofilms comes from the scattering of electrons at the interface between a substrate and the nanofilm, which leads to a broadening of the QWS levels. Though this will smoothen a drop in the lateral
size of the Cooper pairs, our results given in Fig.~\ref{fig2}(d) make it possible to expect that the effect in question will survive to a great extent until the broadening essentially exceeds $5\hbar\omega_D \sim 40\,{\rm meV}$. Here it is worth noting that in experiments with lead nanofilms, the level broadening of QWS can be reduced, in the presence of a crystalline interface, down to $30$-$40\,{\rm meV}$~(at zero temperature and at energies close to $E_F$), see Ref.~\onlinecite{schneider}. This does not exceed $5\,\hbar\omega_D$ mentioned above, which gives quite an optimistic view on the prospects of an experimental observation of the effect in question. We would also like to note that the line width for the perpendicular discrete levels increases with temperature due to electron-phonon scattering. Though such an increase is not very pronounced for temperatures around $T_c$, i.e., it is of about $20\,{\rm meV}$, see Ref.~\onlinecite{schneider}, this will cause an additional temperature-dependent smoothing. So, the optimal choice to probe the BCS-BEC-like crossover in single-crystalline nanofilms is to stay at nearly zero temperatures.

We would also like to highlight strong similarities between our findings and recent angle resolved photoemission results~\cite{lub} for multiband superconducting iron-chalcogenide ${\rm Fe Se_xTe_{1-x}}$. It was found in Ref.~\onlinecite{lub} that the investigated sample of ${\rm Fe Se_xTe_{1-x}}$ does not lay deep in either the BCS or BEC regimes but lays instead in the BCS-BEC crossover domain. As demonstrated above, the same conclusion holds for $4\,{\rm ML}$ and $5\,{\rm ML}$ thick lead nanofilms when the bottom of one of the relevant single-electron subbands approaches $E_F$. Thus, one can in general conclude that the multiband superconductors (both, bulk and with quantum-confined induced subbands, see also Refs.~\onlinecite{shan1} and \onlinecite{inno}) offer new important possibilities towards further understanding the physics of the BCS-BEC crossover in solid-state materials.

\begin{acknowledgments}
This work was supported by the Flemish Science Foundation (FWO-Vl), the Belgian Science Policy (IAP), and the ESF-INSTANS network. A.A.S. thanks A. Bianconi, M. D. Croitoru and A. V. Vagov for useful discussions. A.A.S. acknowledges the hospitality and fruitful interactions with G. C. Strinati, P. Pieri, D. Neilson during his visit to the University of Camerino, supported by the School of Advanced Studies of the University of Camerino.
\end{acknowledgments}


\begin{thebibliography}{99}
\bibitem{note} Typically, $E_F$ only slightly shifts down when increasing
the thickness. However, such a shift becomes more pronounced for
smaller thicknesses, i. e., in the presence of a few single-electron
subbands. In particular, for only one available subband, its bottom
can never cross the Fermi level that simply shifts together with the
subband bottom.
\end{thebibliography}
\end{document}